\documentclass[a4paper,preprint, prl,showpacs]{revtex4}
\usepackage{amsmath,amssymb}
\usepackage{makeidx}
\usepackage{amsfonts}
\usepackage[ansinew]{inputenc}
\usepackage[usenames,dvipsnames]{pstricks}
\usepackage{epsfig,graphicx}

\DeclareMathOperator*{\sgn}{sgn}
\newcommand{\vecSpin}{\mbox{\boldmath$S$}}
\newcommand{\vecSpinhat}{\hat\vecSpin}
\newcommand{\vecPsi}{\mbox{\boldmath$\psi$}}
\newcommand{\vecPsihat}{\hat\vecPsi}

\newcommand{\Spin}{\mbox{$S$}}

\newcommand{\Prob}{\mbox{$P$}}
\newcommand{\Probhat}{\hat{\Prob}}

\newcommand{\Psihat}{\hat\psi}
\newcommand{\nullvec}{\mbox{\boldmath$0$}}
\newcommand{\rme}{\mathrm{e}}
\newcommand{\rmi}{\mathrm{i}}
\newcommand{\rmd}{\mathrm{d}}

\makeindex

\begin{document}

\title{Dynamics of Boolean networks - an exact solution}

\author{Alexander Mozeika}
\author{David Saad}

\affiliation{The Non-linearity and Complexity  Research Group, Aston University, Birmingham B4 7ET, UK.}

\date{\today}

\begin{abstract}
The dynamics of Boolean networks (BN) with quenched disorder and thermal noise is studied via the generating functional method. A general formulation, suitable for BN with any distribution of Boolean functions, is developed. It provides exact solutions and insight into the evolution of order parameters and properties of the stationary states, which are inaccessible via existing methodology. We identify cases where the commonly used annealed approximation is valid and others where it breaks down. Broader links between BN and general Boolean formulas are highlighted.
\end{abstract}


%
\maketitle
%

In his seminal work~\cite{Kauffman} Kauffman introduced a very simple dynamical model of biological gene-regulatory networks. The state of each gene was modeled by an {\small{\sf ON/OFF}} variable, interacting with other genes via a coupling Boolean function which determines the state of a gene at the next time-step. There are $N$ such genes (sites) in the network and each gene is influenced by exactly $k$ other genes from the same network. In Kauffman's approach, the networks are constructed in a random manner by choosing Boolean functions from the set of all $2^{2^k}$ functions of $k$ inputs  and by connecting the inputs of each function to the genes randomly selected from the set $1,..,N$; Boolean functions and connections are fixed for all subsequent time-steps (\emph{quenched} variables). The evolution of a such dynamical system is deterministic and since the number of states is finite ($2^N$) the system is driven to a periodic-orbit \emph{attractor}.

It was argued~\cite{Kauffman} that, despite its simplicity this model, also known as Random Boolean network (RBN) or Kauffman net, is of relevance to the understanding of biological systems and has been studied primarily  for this reason~\cite{RBNbook}. RBN belongs to a larger class of Boolean networks, the N-k model of $N$-variable dynamical systems with a discrete state-space and $k$-variable interactions, that exhibits a rich dynamical behavior~\cite{Aldana,Drossel}. The N-k model is very versatile and has found its use in the modeling of genetic networks~\cite{Nature},  neural networks~\cite{DerridaANN}, social networks~\cite{Moreira} and in many other branches of science~\cite{Aldana,Drossel}.

For over two decades the \emph{annealed approximation}~\cite{Derrida} has proved to be a valuable tool in the analysis of large scale Boolean networks ($N\rightarrow\infty$) as it allows one to predict the time evolution of network activity (proportion of {\small{\sf ON/OFF}} states) and Hamming distance (the difference between the states of two networks of identical topology) order parameters. The latter was used~\cite{Derrida} to predict a phase transition at $k\!=\!2$ in RBN. The main assumption in this method is to ignore the fact that both  Boolean functions types and random connections in a Boolean network are quenched variables and enables one to resample them at each time-step. This allows one to ignore the correlations among input-variables, which simplifies an analytical treatment significantly. It was shown~\cite{DerridaAndW,Hilhorst} that the annealed approximation indeed gives a correct result for the Hamming distance order parameter in RBN, but the broad validity of the annealed approximation to general networks of this type has remained an open problem~\cite{Kesseli}. Remarkably, the annealed approximation provides accurate activity and Hamming distance results for many other Boolean models with quenched disorder but cannot compute correlation functions, used in studying memory effects, due to the repeated resampling at different time steps that makes the various quenched systems indistinguishable. Furthermore, there are models~\cite{RTN} that have very strong memory effects in specific regimes, where the annealed approximation is no longer valid.

In this Letter, we study the dynamics of the N-k model with \emph{quenched} disorder and thermal noise using the generating functional analysis (GFA), an established method for studying physical systems of this type~\cite{dD}; the analysis is general and covers a  large class of recurrent Boolean networks and related models. We show that results for the Hamming distance and network activity obtained via the quenched and annealed approaches, for the N-k model, are identical. In addition, stationary solutions of Hamming distance and two-time autocorrelation function (inaccessible via the annealed approximation) coincide, giving insight into the uniform mapping of states within the basin of attraction onto the stationary states. In the presence of noise, we show that above some noise level the system is always ergodic and explore the possibility of spin-glass phase~\cite{BOOK} below this level. Finally, we show that our theory can be used to study the dynamics of models with strong memory effects.

The model considered is an $N$-variable recurrent Boolean network with the  parallel update rule
\begin{eqnarray}
S_{i}(t\!+\!1)=\alpha_i
(S_{i_1}(t),\ldots,S_{i_k}(t)),\label{def:algorithm}
\end{eqnarray}
where $S_{i}(t)\in\{\!-\!1,1\}$ and $\alpha_i:\{-1,1\}^k\rightarrow\{-1,1\}$ is a Boolean function of exactly $k$ inputs. We  assume
that the thermal noise can flip the output of a function with probability $p$~\cite{Noise}. The function at  site $i$ and time-step $t\!+\!1$ operates in a stochastic manner according to the microscopic law
\begin{eqnarray}
&&\Prob_{\alpha_i}(S_{i}(t\!+\!1)\vert S_{i_1}(t),..,S_{i_k}(t))\label{eq:micro}\\
&&=\frac{\rme^{\beta S_{i}(t\!+\!1)\alpha_i(S_{i_1}(t),..,S_{i_k}(t))}}{2\cosh\beta\alpha_i(S_{i_1}(t),..,S_{i_k}(t))}\nonumber
\end{eqnarray}
where the inverse temperature $\beta\!=\!1/T$ relates to the noise parameter $p$ via $\tanh\beta\!=\!1\!-\!2p$. The function-output $S_i(t\!+\!1)$ is completely random/deterministic when $\beta\!\rightarrow\!0/\infty$, respectively.  Given the state of the network $\vecSpin(t)\in\{\!-\!1,1\}^N$  at time $t$  the functions at time $t\!+\!1$ are independent of each other. This suggests that the probability of the microscopic path $\vecSpin(0)\rightarrow\!\cdots\!\rightarrow\vecSpin(t_{max})$ is a product of (\ref{eq:micro}) over  sites and time steps. The joint probability of microscopic states in two systems of identical topology but subject to different thermal noise is
\begin{eqnarray}
\Prob[\{\vecSpin(t)\};&\phantom{=}&\hspace*{-2em}\{\vecSpinhat(t)\}]\!=\!\Prob(\vecSpin(0),\vecSpinhat(0)) \label{eq:PathProb}\\ \times \prod_{t=0}^{t_{max}-1} &\phantom{=}&\hspace*{-2em} \Prob(\vecSpin(t\!+\!1)\vert\vecSpin(t))P(\vecSpinhat(t\!+\!1)\vert\vecSpinhat(t)) \mbox{~where,}\nonumber
\end{eqnarray}
$\Prob(\vecSpin(t\!+\!1)\vert\vecSpin(t))\!=\!\prod_{i\!=\!1}^N\Prob_{\alpha_i}(\Spin_{i}(t\!+\!1)\vert \Spin_{i_1}(t),..,\Spin_{i_k}(t))$.

The quenched disorder in our model arises from the random sampling of connections and Boolean functions generated by selecting the $i$-th function and sampling exactly $k$ indices, $\{i_1,..,i_k\}$, uniformly from the set of all possible indices. Boolean functions $\{\alpha_i\}$ are sampled randomly and independently from the set $G$ of $k$-ary Boolean functions. To analyze the typical properties of the system via the generating functional method one defines
\begin{eqnarray}
\Gamma[\vecPsi;\vecPsihat]&=&\left\langle\rme^{-\rmi\sum_{t,i}\{\psi_i(t) S_{i}(t)+\Psihat_i(t) \hat{S}_{i}(t)\}}\right\rangle~,\label{eq:GF}
\end{eqnarray}	
where $\langle\ldots\rangle$ denotes the average generated by (\ref{eq:PathProb}). The generating function (\ref{eq:GF}) is used to compute moments of (\ref{eq:PathProb}) by taking partial derivatives with respect to the generating fields $\{\psi_i(t),\Psihat_j(s)\}$, e.g. $\langle S_i(t)\hat{S}_j(s)\rangle=-\lim_{\vecPsi,\vecPsihat\rightarrow\nullvec} \frac{\partial^2}{\partial_{\psi_i(t)}\partial_{\hat{\psi}_j(s)}}\Gamma[\vecPsi;\vecPsihat]$. We assume that the system becomes self-averaging for $N\rightarrow\infty$~\cite{dD} and  compute $\overline{\Gamma[\vecPsi;\vecPsihat]}$, where $\overline{\cdots}$ is the disorder average; this gives rise to the macroscopic observables
\begin{eqnarray}
&&m(t)\!=\!\frac{1}{N}\sum_{i=1}^N\overline{\langle S_i(t)\rangle},\; C(t,\! s)\!=\!\frac{1}{N}\!\!\sum_{i=1}^N\overline{\langle S_i(t)S_i(s)\rangle}\label{def:observ}\\
&&C_{12}(t)\!=\!\frac{1}{N}\!\sum_{i=1}^N\overline{\langle S_i(t)\hat{S}_i(t)\rangle}\nonumber
\end{eqnarray}
where $m(t)$ is the network activity (or magnetization~\cite{MinRBN}), $C(t,\! s)$ is the correlation between two states of the same network  and $C_{12}(t)$ (related to the Hamming distance $d(t)$ via $d(t)=\frac{1}{2}(1\!-\!C_{12}(t))$) is the overlap between two copies of the same network.

Averaging (\ref{eq:GF}) over the disorder~\cite{longpaper} leads to the saddle-point integral $\overline{\Gamma[\ldots]}\!=\!\int\{\rmd P\rmd\hat P\}\rme^{N\Psi[P,\hat P]}$ where
\begin{equation}\Psi\!=\!\rmi\sum_{ \vecSpin,\vecSpinhat}\!\Probhat(\vecSpin,\vecSpinhat)\Prob(\vecSpin,\vecSpinhat)\label{eq:saddle}
\!+\! \log \!\sum_{\vecSpin,\vecSpinhat}\!\Prob( \vecSpin,\vecSpinhat)\rme^{-\rmi\hat{P}(\vecSpin,\vecSpinhat)}~.\end{equation}
For $N\!\rightarrow\!\infty$ the averaged generating functional is dominated by the extremum of $\Psi$. Functional variation with respect to the order parameters $\hat{P}(\vecSpin,\vecSpinhat)$ provides the saddle-point equation
%
\begin{widetext}
\begin{eqnarray}
&&\Prob( \vecSpin,\vecSpinhat)\!=\!\Prob(S(0),\hat{S}(0))\!\!\!\!\sum_{\{\vecSpin_j,\vecSpinhat_j\}}\!\prod_{j=1}^{k}\!\left[\Prob(\vecSpin_j,\vecSpinhat_j)\right]\nonumber\\
&&\times\left\langle \prod_{t=0}^{t_{max}-1}\!\!\!\!\Prob_{\alpha}(S(t\!+\!1)\vert S_{1}(t),..,S_{k}(t))\Prob_{\alpha}(\hat{S}(t\!+\!1)\vert \hat{S}_{1}(t),..,\hat{S}_{k}(t))\right\rangle_\alpha.\label{eq:M}
\end{eqnarray}
\end{widetext}
%
The physical meaning of (\ref{eq:M}) relates to the average joint probability of single-spin trajectories $\vecSpin$ and $\vecSpinhat$ in the two systems $\Prob(\vecSpin,\vecSpinhat)\!=\!\lim_{N\rightarrow\infty}\frac{1}{N}\sum_{i=1}^N\overline{\langle\delta[\vecSpin;\vecSpin_i]\,\delta[\vecSpinhat;\vecSpinhat_i]\rangle}$, while the conjugate order parameter $\Probhat(\vecSpin,\vecSpinhat)$ is a constant. Equation (\ref{eq:M}) can be used to compute the macroscopic observables (\ref{def:observ}), which evolve in time as follows below, denoting $S=(S_1,\ldots,S_k)$ and where the magnetization $\hat m(t)$ is computed by (\ref{eq:m})
\begin{widetext}
\begin{eqnarray}
&&m(t\!+\!1)=f_\alpha(m(t))\!=\!\tanh(\beta)\sum_{S}\prod_{j=1}^{k}\left[\frac{1\!+\! S_jm(t)}{2}\right]\left\langle  \alpha(S) \right\rangle_\alpha\label{eq:m}\\
&&C(t\!+\!1,s\!+\!1)\!=\!F_\alpha(m(t),m(s),C(t,s))\nonumber\\
&&\!=\!\tanh^2(\beta)\sum_{S,\hat{S}}\!\prod_{j=1}^{k}\!\left[\!\frac{1\!+\! S_j m(t)\!+\!\hat{S}_j m(s) \!+\! \!S_j\hat  S_j C(t,s)}{4}\right]\langle  \alpha(S)\alpha(\hat{S})\rangle_\alpha\label{eq:Corr}\\
&&C_{12}(t\!+\!1)=F_\alpha(m(t),\hat m(t),C_{12}(t)),\label{eq:overlap}
\end{eqnarray}
\end{widetext}
%
Results for the order parameters (\ref{eq:m})-(\ref{eq:overlap}), in combination with (\ref{eq:M}), suggest that the evolution of all many-time single-site correlation functions is driven by the magnetization $m(t)$. A similar scenario was observed in recurrent asymmetric neural networks~\cite{MimuraAndCoolen}, defined on similar topology due to similarity in the equations for $m(t)$ and $C(t,s)$. This is not surprising since asymmetric neural network is a special case of the N-k model when only \emph{linear threshold Boolean functions} are used. Furthermore, for the stationary  solution $m\!=\!f_\alpha(m)$ ($m\!=\!\lim_{t\rightarrow\infty} m(t)$) the solutions of $q\!=\!F_\alpha(m,m,q)$ (here $q\!=\!\lim_{t\rightarrow\infty}\lim_{\tau\rightarrow\infty}C(t\!+\!\tau,\tau)$ is the Edwards-Anderson order parameter, used in disordered systems~\cite{BOOK} to detect the spin glass phase where $m\!=\!0$ and $q\!\neq\!0$) and $C_{12}\!=\!F_\alpha(m,m,C_{12})$ are identical. This suggests that there is only one average distance $\frac{1}{2}(1\!-\!q)$ on the attractor~\cite{adnn} and that all points in the basin of attraction uniformly cover the stationary states.

The annealed model, where connectivities and Boolean functions change at each time step (\ref{def:algorithm}) provides \emph{identical results} for $m$ and $C_{12}$ to those of (\ref{eq:m}) and (\ref{eq:overlap})~\cite{Kesseli}. However, the annealed correlation function $C(t,s)\!=\!m(t)m(s)$, where $t\!>\!s$, is the solution of (\ref{eq:Corr}) \emph{only when networks are constructed from a single function type.}

The annealed result~\cite{Derrida} for RBN can be easily recovered from  equations (\ref{eq:m})-(\ref{eq:overlap}) using the property $\langle \alpha(S)\rangle_\alpha\!=\!0$ for all $S\!\in\!\{-1,1\}^k$ and $\langle \alpha(S)\alpha(\hat S)\rangle_\alpha\!=\!0$ $,\forall S\!\neq\!\hat S$ where the $\alpha$ average is taken over all Boolean functions with equal weight. In this case, the magnetization $m(t)\!=\!0$ for all $t\!>\!0$ and  $q=\tanh^2(\beta)(\frac{1\!+\!q}{2})^k$, corresponding to the stationary solution of (\ref{eq:Corr}), has one stable solution $q\!\neq\!0$ for all finite $\beta$ and $k$. For $\beta\!\rightarrow\!\infty$ (no noise), a transition is observed from one stable solution $q=1$ for $k\!\leq\!2$ to two solutions $q\!=\!1$ (unstable) and $q\!\neq\!0$ (stable) for $k\!>\!2$~\cite{Derrida}.

The unordered paramagnetic  phase $m\!=\!0$ is a fixed point of~(\ref{eq:m}) only when  $\sum_S\left\langle \alpha(S)\!\right\rangle_\alpha\!=\!0$. This is a stable and unique solution of (\ref{eq:m}) when $\tanh\beta\!<\!\left\{2^{k\!-\!1}/k\binom{k\!-\!1}{(k\!-\!1)/2}; 2^{k\!-\!2}/(k\!-\!1)\binom{k\!-\!2}{(k\!-\!2)/2}\right\}\!\equiv \!b(k)$ for $k$ odd and even respectively. To prove this~\cite{longpaper} we first find a Boolean function $\chi$ such that $f_\chi(m)\!\geq\! f_\alpha(m)$ when $m\!\in\![0,1)$ and $f_\alpha(m)\!\geq\! f_\chi(m)$ when $m\!\in\!(-1,0]$; any function from the set $\chi(S)\!=\!\sgn[\sum_{j=1}^k S_j]\!+\!\delta[0;\sum_{j=1}^k S_j]\gamma(S)$, where $\gamma(S)\in\{\!-\!1,1\}$ and such that $\sum_S\delta[0;\sum_{j=1}^k S_j]\gamma(S)\!=\!0$ \footnote{We use the convention $\sgn[0]\!=\!0$ throughout this Letter.} satisfies these properties. Secondly, we show that  $m\!>\!f_\chi(m)$ when  $m\!\in\!(0,1)$ and $f_\chi(m)\!>\!m$ when $m\!\in\!(-1,0)$ ($f_\chi(0)\!=\!0$) for $\tanh\beta\!<\!b(k)$. Thus, the ordered (ferromagnetic) phase $m\!\neq\!0$  is a fixed point of (\ref{eq:m}) (if at all) \emph{only} for values of $\beta$ and $k$ which satisfy $\tanh\beta\!>\!b(k)$. Similar results, for odd $k$ only, have been conjectured using the annealed approximation and multiplexing techniques~\cite{Peixoto}.

For $\lim_{t\rightarrow\infty}m(t)\!=\!m$, $q\!=\!0$ is a fixed point of (\ref{eq:Corr}) iff $\left\langle \{\sum_S\alpha(S)\}^2\right\rangle_\alpha\!=\!0$ which occurs only for \emph{balanced Boolean functions}, with an equal number of $\!\pm\!1$ in the output. By similar argument to the one used in the previous paragraph we show~\cite{longpaper} that for $m\!=\!0$ the point $q\!=\!0$ is a unique stable solution of (\ref{eq:Corr}) when $\tanh^2\beta\!<\!b(k)$. The $\alpha$-averages in equations (\ref{eq:m})-(\ref{eq:Corr}) can be computed for a uniform distribution over all balanced Boolean functions to obtain $m(t)\!=\!0$ for all $t\!>\!0$, which implies $q\!=\!\tanh^2(\beta)\left((\frac{1\!+\!q}{2})^k(1\!+\!\frac{1}{2^k\!-\!1})\!-\!\frac{1}{2^k\!-\!1}\right)$. The latter has only one $q\!=\!0$ trivial solution for any finite $\beta$ and develops a second $q\!=\!1$ solution only for $\beta\!\rightarrow\!\infty$. Thus, the case of $m\!=\!0$, $q\!\neq\!0$ and finite  $\beta$ occurs only (if at all) when $\tanh^2\beta\!>\!b(k)$ \emph{and} for non-uniform distributions over the balanced Boolean functions.

The upper bound $b(k)$ computed here for $k$ odd is identical to the one computed for noisy Boolean formulas~\cite{Evans:MTNK}. This is since each site $i$ at time $t$ in our model can be associated with the output $S_i(t)$ of a $k$-ary Boolean formula of depth $t$  which computes a function of the associated initial states (a subset of $\{S_i(0)\}$)~\cite{DerridaAndW}. In the presence of noise, a formula of considerable depth (large $t$) loses all input information for  $\tanh\beta\!<\!b(k)$ and odd $k$~\cite{Evans:MTNK}. This suggests that the upper bound $b(k)$, for odd  $k$, is more general and is valid for transitions at \emph{all} $m$ values identifying the point where stationary states depend on the initial states and ergodicity breaks.  For $k$ even such \emph{general} threshold is not yet known.

%
In model~(\ref{def:algorithm}) the state of site $i$ at time $t$ depends on its states at previous times only indirectly. In the limit of  $N\!\rightarrow\!\infty$ these dependencies become weak and equation~(\ref{eq:M}) factorizes; this enables one to calculate the observables of interest (\ref{eq:m})-(\ref{eq:overlap}). However, in a broad family of models~\cite{Revers,DeSales} the state of a site $i$ at a time $t\!+\!1$ depends directly on its state at time $t$. An exemplar model with strong memory effects used to construct a model of cell-cycle regulatory network ($N\!=\!11$) of budding yeast~\cite{Li} is of the form
\begin{equation}
S_i(t\!+\!1)\!=\!\sgn[h_i(t)\!-\!2h]\!+\!S_i(t)\delta[h_i(t);2h]\label{eq:process},
\end{equation}
where $h_i(t)\!=\!\sum_{j=1}^k\xi_{i_j}(1\!+\!S_{i_j}(t))$ and $\xi_{i_j}\!\in\!\{\!-\!1,1\}$. Mean-field theory ($N\!\rightarrow\!\infty$) was derived~\cite{RTN} using the annealed approximation in a variant of this model, where the interactions $\{\xi_j\}$ were randomly distributed $\Prob(\xi_j\!=\!\pm1)\!=\!1/2$. Significant discrepancies between the theory and simulation results has been pointed out~\cite{RTN} for  integer $h$ values (in this case it is possible that $2h\!=\!h_i(t)$), which was attributed to the presence of strong memory effects. Refinements of the annealed approximation method improved the results obtained only slightly~\cite{Heckel,Zanudo} but break down in most of the parameter space.

This model (\ref{eq:process}) can be easily incorporated into our theoretical framework. The result of the GFA (\ref{eq:M}) for this process (with thermal noise) can be obtained by replacing the  average $\langle\cdots\rangle_\alpha$ by $\langle\cdots\rangle_\xi$ and the probability function $\Prob_{\alpha}(S(t\!+\!1)\vert S_{1}(t),..,S_{k}(t))$ by
\begin{eqnarray}
&&\Prob_{\xi}(S(t\!+\!1)\vert S(t);S_1(t),..,S_k(t))=\label{def:ProbRTN}\\
&&\frac{\rme^{\beta S(t\!+\!1)\{\sgn[h(t)\!-\!2h]\!+\!S(t)\delta[h(t);2h]\}}}{2\cosh\beta\{\sgn[h(t)\!-\!2h]\!+\!S(t)\delta[h(t);2h]\}},\nonumber
\end{eqnarray}
where $h(t)\!=\!\sum_{j=1}^k\xi_{j}(1\!+\!S_{j}(t))$.
\begin{figure}[t]
\vspace*{0mm} \hspace*{-0mm} \setlength{\unitlength}{0.27mm}
\begin{picture}(350,210)
\put(157,110){\includegraphics[height=100\unitlength,width=160\unitlength]{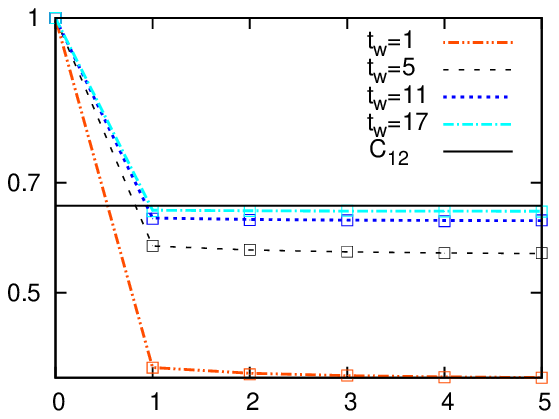}}
\put(0,110){\includegraphics[height=100\unitlength,width=160\unitlength]{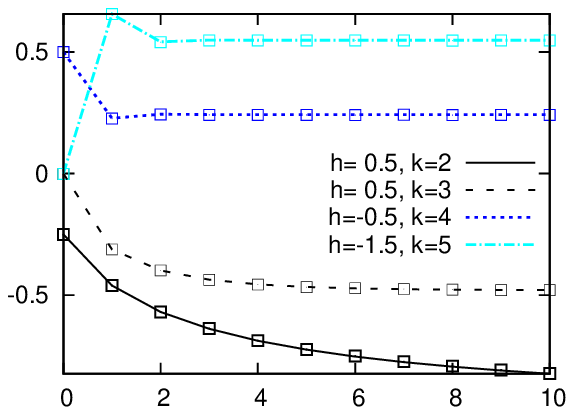}}
\put(159,0){\includegraphics[height=100\unitlength,width=160\unitlength]{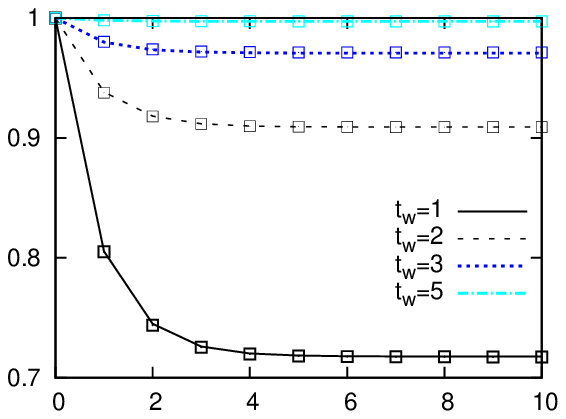}}
\put(245,-7){$t$} 
\put(0,0){\includegraphics[height=100\unitlength,width=160\unitlength]{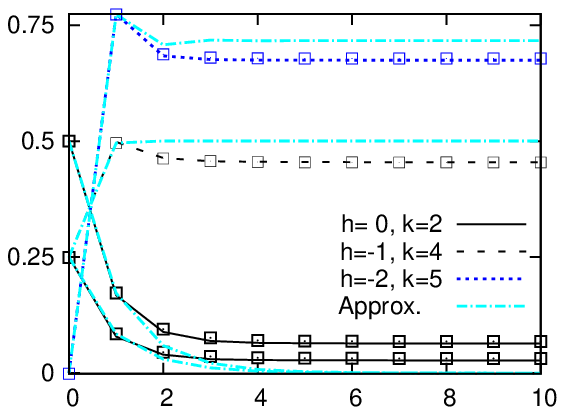}}
\put(-3,175){$m$} \put(-3,65){$m$}
\put(85,-7){$t$} \put(160,175){$C$} \put(160,65){$C$}
\put(130,190){$(a)$}\put(290,190){$(b)$}\put(130,80){$(c)$}\put(290,80){$(d)$}
\end{picture}
 \vspace*{0mm}
\caption{(Color online). Evolution of the magnetization ($m\equiv m(t)$) and correlation ($C\equiv C(t\!+\!t_w,t_w)$) functions with time $t$ is governed by (\ref{eq:process}). Theoretical results (lines) are plotted against the results of MC simulations (symbols) with $N\!=\!10^5$. Each MC data-point is averaged over 10 runs. Error bars are smaller than symbol size. Top: Evolution of $m$ (a) and $C$ (b) for $h\!\in\!\mathbb{R}$. In (b) we plot $C$ for $h\!=\!0.5$ and $k\!=\!3$. Bottom: Evolution of $m$ (c) and $C$ (d) for $h\!\in\!\mathbb{Z}$. In (d) we plot $C$ for $h\!=\!0$ and $k\!=\!2$.\label{fig:0} \vspace*{-0.5cm}
}
\end{figure}
In the case of $h\!\in\!\mathbb{R}$, the probability function (\ref{def:ProbRTN}) is independent of $S(t)$  and equations (\ref{eq:m})-(\ref{eq:overlap}) have the same structure as model~(\ref{eq:process}): the $\alpha$-averages $\langle \alpha(S)\rangle_\alpha$ and $\langle \alpha(S)\alpha(\hat S)\rangle_\alpha$ are replaced by the averages $\langle\sgn[h(t)\!-\!2h]\rangle_\xi$ and $\langle\sgn[h(t)\!-\!2h]\sgn[\hat h(t)\!-\!2h]\rangle_\xi$ respectively. The equation for $m(t)$ recovers the annealed approximation result~\cite{RTN} (using the relation $b(t)\!=\!(1\!+\!m(t))/2$). In Fig.~\ref{fig:0} (a,b), we plot our analytical predictions for the evolution of $m(t)$ and $C(t\!+\!t_w,t_w)$ against the results of Monte Carlo (MC) simulation which use (\ref{eq:process}). The correlation function $C(t\!+\!t_w,t_w)$, in the limit of  $t\!\rightarrow\!\infty,t_w\!\rightarrow\!\infty$, approaches the stationary solution of the overlap function (\ref{eq:overlap})  as predicted (Fig.~\ref{fig:0}(b)).

The situation is very different when $h\!\in\!\mathbb{Z}$. Then the magnetization $m(t)\!=\!\sum_{\vecSpin}\Prob( \vecSpin)S(t)$ where $\Prob(\vecSpin)$ is a marginal of (\ref{eq:M}) with $P_\alpha\!\rightarrow\! P_\xi$, is no longer closed as in (\ref{eq:m}), but depends on $2^{t\!-\!1}\!-\!1$ macroscopic observables (all magnetizations, all multi-time correlations). Thus the number of macroscopic observables that determine the value of  $m(t)$, or any other function computed from (\ref{eq:M}), grows exponentially with time. Annealed approximation results~\cite{RTN} for this model when $h\!\in\!\mathbb{Z}$ are only exact up to $t\!<\!2$ time steps (the equation for $b(1)\!=\!(1\!+\!m(1))/2$ in our approach and in~\cite{RTN} are identical) and deviate significantly from the exact solution at later times (Fig.~\ref{fig:0}(c)). A typical evolution of the correlation function $C(t\!+\!t_w,t_w)$ in the system (\ref{eq:process}) when $h\in\mathbb{Z}$ is shown in Fig.~\ref{fig:0}(d).

As BNs are instrumental for our understanding of biological and other complex networks, and are directly linked to general Boolean formulas there is a need to develop exact tools of greater flexibility that cope with complex networks of variable Boolean functions with strong memory effects and emerging correlations. This Letter is the first step in this direction.

\begin{acknowledgments}
 Support by the Leverhulme trust (grant F/00 250/H) is acknowledged.
\end{acknowledgments}


\begin{thebibliography}{26}
\expandafter\ifx\csname natexlab\endcsname\relax\def\natexlab#1{#1}\fi
\expandafter\ifx\csname bibnamefont\endcsname\relax
  \def\bibnamefont#1{#1}\fi
\expandafter\ifx\csname bibfnamefont\endcsname\relax
  \def\bibfnamefont#1{#1}\fi
\expandafter\ifx\csname citenamefont\endcsname\relax
  \def\citenamefont#1{#1}\fi
\expandafter\ifx\csname url\endcsname\relax
  \def\url#1{\texttt{#1}}\fi
\expandafter\ifx\csname urlprefix\endcsname\relax\def\urlprefix{URL }\fi
\providecommand{\bibinfo}[2]{#2}
\providecommand{\eprint}[2][]{\url{#2}}

\bibitem[{\citenamefont{{Kauffman}}(1969{\natexlab{a}})}]{Kauffman}
\bibinfo{author}{\bibfnamefont{S.~A.} \bibnamefont{{Kauffman}}},
  \bibinfo{journal}{J. Theor. Biol.} \textbf{\bibinfo{volume}{22}},
  \bibinfo{pages}{437} (\bibinfo{year}{1969}{\natexlab{a}}).

\bibitem[{\citenamefont{Kauffman}(1993)}]{RBNbook}
\bibinfo{author}{\bibfnamefont{S.}~\bibnamefont{Kauffman}},
  \emph{\bibinfo{title}{The Origins of Order}} (\bibinfo{publisher}{Oxford
  University Press}, \bibinfo{address}{New York}, \bibinfo{year}{1993}).

\bibitem[{\citenamefont{Aldana et~al.}(2003)\citenamefont{Aldana, Coppersmith,
  and Kadanoff}}]{Aldana}
\bibinfo{author}{\bibfnamefont{M.}~\bibnamefont{Aldana}},
  \bibinfo{author}{\bibfnamefont{S.}~\bibnamefont{Coppersmith}},
  \bibnamefont{and} \bibinfo{author}{\bibfnamefont{L.~P.}
  \bibnamefont{Kadanoff}}, \emph{\bibinfo{title}{Perspectives and Problems in
  Nonlinear Science. A Celebratory Volume in Honor of Lawrence Sirovich.}}
  (\bibinfo{publisher}{Springer}, \bibinfo{address}{New York},
  \bibinfo{year}{2003}), chap. \bibinfo{chapter}{Boolean dynamics with random
  couplings}, pp. \bibinfo{pages}{23--89}.

\bibitem[{\citenamefont{{Drossel}}(2008)}]{Drossel}
\bibinfo{author}{\bibfnamefont{B.}~\bibnamefont{{Drossel}}},
  \emph{\bibinfo{title}{Random Boolean networks}} (\bibinfo{publisher}{Wiley},
  \bibinfo{address}{Weinheim}, \bibinfo{year}{2008}), vol.~\bibinfo{volume}{1}
  of \emph{\bibinfo{series}{Reviews of Nonlinear Dynamics and Complexity}},
  chap.~\bibinfo{chapter}{3}, pp. \bibinfo{pages}{69--96}.

\bibitem[{\citenamefont{{Kauffman}}(1969{\natexlab{b}})}]{Nature}
\bibinfo{author}{\bibfnamefont{S.}~\bibnamefont{{Kauffman}}},
  \bibinfo{journal}{Nature} \textbf{\bibinfo{volume}{224}},
  \bibinfo{pages}{177} (\bibinfo{year}{1969}{\natexlab{b}}).

\bibitem[{\citenamefont{{Derrida} et~al.}(1987)\citenamefont{{Derrida},
  {Gardner}, and {Zippelius}}}]{DerridaANN}
\bibinfo{author}{\bibfnamefont{B.}~\bibnamefont{{Derrida}}},
  \bibinfo{author}{\bibfnamefont{E.}~\bibnamefont{{Gardner}}},
  \bibnamefont{and}
  \bibinfo{author}{\bibfnamefont{A.}~\bibnamefont{{Zippelius}}},
  \bibinfo{journal}{Europhys. Lett.} \textbf{\bibinfo{volume}{4}},
  \bibinfo{pages}{167} (\bibinfo{year}{1987}).

\bibitem[{\citenamefont{{Moreira} et~al.}(2004)\citenamefont{{Moreira},
  {Mathur}, {Diermeier}, and {Amaral}}}]{Moreira}
\bibinfo{author}{\bibfnamefont{A.~A.} \bibnamefont{{Moreira}}},
  \bibinfo{author}{\bibfnamefont{A.}~\bibnamefont{{Mathur}}},
  \bibinfo{author}{\bibfnamefont{D.}~\bibnamefont{{Diermeier}}},
  \bibnamefont{and} \bibinfo{author}{\bibfnamefont{L.~A.~N.}
  \bibnamefont{{Amaral}}}, \bibinfo{journal}{Proc. Nat. Acad. Sci. U.S.A.}
  \textbf{\bibinfo{volume}{101}}, \bibinfo{pages}{12085}
  (\bibinfo{year}{2004}).

\bibitem[{\citenamefont{{Derrida} and {Pomeau}}(1986)}]{Derrida}
\bibinfo{author}{\bibfnamefont{B.}~\bibnamefont{{Derrida}}} \bibnamefont{and}
  \bibinfo{author}{\bibfnamefont{Y.}~\bibnamefont{{Pomeau}}},
  \bibinfo{journal}{Europhys. Lett.} \textbf{\bibinfo{volume}{1}},
  \bibinfo{pages}{45} (\bibinfo{year}{1986}).

\bibitem[{\citenamefont{{Derrida} and {Weisbuch}}(1986)}]{DerridaAndW}
\bibinfo{author}{\bibfnamefont{B.}~\bibnamefont{{Derrida}}} \bibnamefont{and}
  \bibinfo{author}{\bibfnamefont{G.}~\bibnamefont{{Weisbuch}}},
  \bibinfo{journal}{J. Phys.} \textbf{\bibinfo{volume}{47}},
  \bibinfo{pages}{1297} (\bibinfo{year}{1986}).

\bibitem[{\citenamefont{Hilhorst and Nijmeijer}(1987)}]{Hilhorst}
\bibinfo{author}{\bibfnamefont{H.~J.} \bibnamefont{Hilhorst}} \bibnamefont{and}
  \bibinfo{author}{\bibfnamefont{M.}~\bibnamefont{Nijmeijer}},
  \bibinfo{journal}{J. Phys.} \textbf{\bibinfo{volume}{48}},
  \bibinfo{pages}{185} (\bibinfo{year}{1987}).

\bibitem[{\citenamefont{Kesseli et~al.}(2006)\citenamefont{Kesseli, R\"am\"o,
  and Yli-Harja}}]{Kesseli}
\bibinfo{author}{\bibfnamefont{J.}~\bibnamefont{Kesseli}},
  \bibinfo{author}{\bibfnamefont{P.}~\bibnamefont{R\"am\"o}}, \bibnamefont{and}
  \bibinfo{author}{\bibfnamefont{O.}~\bibnamefont{Yli-Harja}},
  \bibinfo{journal}{Phys. Rev. E.} \textbf{\bibinfo{volume}{74}},
  \bibinfo{pages}{046104} (\bibinfo{year}{2006}).

\bibitem[{\citenamefont{{Szejka} et~al.}(2008)\citenamefont{{Szejka},
  {Mihaljev}, and {Drossel}}}]{RTN}
\bibinfo{author}{\bibfnamefont{A.}~\bibnamefont{{Szejka}}},
  \bibinfo{author}{\bibfnamefont{T.}~\bibnamefont{{Mihaljev}}},
  \bibnamefont{and}
  \bibinfo{author}{\bibfnamefont{B.}~\bibnamefont{{Drossel}}},
  \bibinfo{journal}{New J. Phys.} \textbf{\bibinfo{volume}{10}},
  \bibinfo{pages}{063009} (\bibinfo{year}{2008}).

\bibitem[{\citenamefont{De~Dominicis}(1978)}]{dD}
\bibinfo{author}{\bibfnamefont{C.}~\bibnamefont{De~Dominicis}},
  \bibinfo{journal}{Phys. Rev. B.} \textbf{\bibinfo{volume}{18}},
  \bibinfo{pages}{4913} (\bibinfo{year}{1978}).

\bibitem[{\citenamefont{Mezard et~al.}(1987)\citenamefont{Mezard, Parisi, and
  Virasoro}}]{BOOK}
\bibinfo{author}{\bibfnamefont{M.}~\bibnamefont{Mezard}},
  \bibinfo{author}{\bibfnamefont{G.}~\bibnamefont{Parisi}}, \bibnamefont{and}
  \bibinfo{author}{\bibfnamefont{M.~A.} \bibnamefont{Virasoro}},
  \emph{\bibinfo{title}{Spin glass theory and beyond}}
  (\bibinfo{publisher}{World Scientific}, \bibinfo{address}{Singapore},
  \bibinfo{year}{1987}).

\bibitem[{\citenamefont{{Peixoto} and {Drossel}}(2009)}]{Noise}
\bibinfo{author}{\bibfnamefont{T.~P.} \bibnamefont{{Peixoto}}}
  \bibnamefont{and}
  \bibinfo{author}{\bibfnamefont{B.}~\bibnamefont{{Drossel}}},
  \bibinfo{journal}{Phys. Rev. E.} \textbf{\bibinfo{volume}{79}},
  \bibinfo{pages}{036108} (\bibinfo{year}{2009}).

\bibitem[{\citenamefont{{Derrida} and {Flyvbjerg}}(1987)}]{MinRBN}
\bibinfo{author}{\bibfnamefont{B.}~\bibnamefont{{Derrida}}} \bibnamefont{and}
  \bibinfo{author}{\bibfnamefont{H.}~\bibnamefont{{Flyvbjerg}}},
  \bibinfo{journal}{J. Phys. A: Math. Gen.} \textbf{\bibinfo{volume}{20}},
  \bibinfo{pages}{L1107} (\bibinfo{year}{1987}).

\bibitem[{\citenamefont{Mozeika and Saad}(2010)}]{longpaper}
\bibinfo{author}{\bibfnamefont{A.}~\bibnamefont{Mozeika}} \bibnamefont{and}
  \bibinfo{author}{\bibfnamefont{D.}~\bibnamefont{Saad}}
  (\bibinfo{year}{2010}), \bibinfo{note}{in preparation}.

\bibitem[{\citenamefont{Mimura and Coolen}(2009)}]{MimuraAndCoolen}
\bibinfo{author}{\bibfnamefont{K.}~\bibnamefont{Mimura}} \bibnamefont{and}
  \bibinfo{author}{\bibfnamefont{A.~C.~C.} \bibnamefont{Coolen}},
  \bibinfo{journal}{J. Phys. A: Math. Theor.} \textbf{\bibinfo{volume}{42}},
  \bibinfo{pages}{415001} (\bibinfo{year}{2009}).

\bibitem[{\citenamefont{{Kree} and {Zippelius}}(1987)}]{adnn}
\bibinfo{author}{\bibfnamefont{R.}~\bibnamefont{{Kree}}} \bibnamefont{and}
  \bibinfo{author}{\bibfnamefont{A.}~\bibnamefont{{Zippelius}}},
  \bibinfo{journal}{Phys. Rev. A.} \textbf{\bibinfo{volume}{36}},
  \bibinfo{pages}{4421} (\bibinfo{year}{1987}).

\bibitem[{\citenamefont{{Peixoto}}(2010)}]{Peixoto}
\bibinfo{author}{\bibfnamefont{T.}~\bibnamefont{{Peixoto}}},
  \bibinfo{journal}{Phys. Rev. Lett.} \textbf{\bibinfo{volume}{104}},
  \bibinfo{pages}{048701} (\bibinfo{year}{2010}).

\bibitem[{\citenamefont{Evans and Schulman}(2003)}]{Evans:MTNK}
\bibinfo{author}{\bibfnamefont{W.}~\bibnamefont{Evans}} \bibnamefont{and}
  \bibinfo{author}{\bibfnamefont{L.}~\bibnamefont{Schulman}},
  \bibinfo{journal}{IEEE Trans. Inf. Theory} \textbf{\bibinfo{volume}{49}},
  \bibinfo{pages}{3094} (\bibinfo{year}{2003}).

\bibitem[{\citenamefont{Coppersmith et~al.}(2001)\citenamefont{Coppersmith,
  Kadanoff, and Zhang}}]{Revers}
\bibinfo{author}{\bibfnamefont{S.~N.} \bibnamefont{Coppersmith}},
  \bibinfo{author}{\bibfnamefont{L.~P.} \bibnamefont{Kadanoff}},
  \bibnamefont{and} \bibinfo{author}{\bibfnamefont{Z.}~\bibnamefont{Zhang}},
  \bibinfo{journal}{Physica D} \textbf{\bibinfo{volume}{149}},
  \bibinfo{pages}{11 } (\bibinfo{year}{2001}).

\bibitem[{\citenamefont{de~Sales et~al.}(1997)\citenamefont{de~Sales, Martins,
  and Stariolo}}]{DeSales}
\bibinfo{author}{\bibfnamefont{J.~A.} \bibnamefont{de~Sales}},
  \bibinfo{author}{\bibfnamefont{M.~L.} \bibnamefont{Martins}},
  \bibnamefont{and} \bibinfo{author}{\bibfnamefont{D.~A.}
  \bibnamefont{Stariolo}}, \bibinfo{journal}{Phys. Rev. E.}
  \textbf{\bibinfo{volume}{55}}, \bibinfo{pages}{3262} (\bibinfo{year}{1997}).

\bibitem[{\citenamefont{Li et~al.}(2004)\citenamefont{Li, Long, Lu, Ouyang, and
  Tang}}]{Li}
\bibinfo{author}{\bibfnamefont{F.}~\bibnamefont{Li}},
  \bibinfo{author}{\bibfnamefont{T.}~\bibnamefont{Long}},
  \bibinfo{author}{\bibfnamefont{Y.}~\bibnamefont{Lu}},
  \bibinfo{author}{\bibfnamefont{Q.}~\bibnamefont{Ouyang}}, \bibnamefont{and}
  \bibinfo{author}{\bibfnamefont{C.}~\bibnamefont{Tang}},
  \bibinfo{journal}{Proc. Nat. Acad. Sci. U.S.A.}
  \textbf{\bibinfo{volume}{101}}, \bibinfo{pages}{4781} (\bibinfo{year}{2004}).

\bibitem[{\citenamefont{Heckel et~al.}(2010)\citenamefont{Heckel, Schober, and
  Bossert}}]{Heckel}
\bibinfo{author}{\bibfnamefont{R.}~\bibnamefont{Heckel}},
  \bibinfo{author}{\bibfnamefont{S.}~\bibnamefont{Schober}}, \bibnamefont{and}
  \bibinfo{author}{\bibfnamefont{M.}~\bibnamefont{Bossert}}, in
  \emph{\bibinfo{booktitle}{Source and Channel Coding (SCC), 2010 International
  ITG Conference on}} (\bibinfo{year}{2010}), pp. \bibinfo{pages}{1 --6}.

\bibitem[{\citenamefont{{Za{\~n}udo} et~al.}(2010)\citenamefont{{Za{\~n}udo},
  {Aldana}, and {Mart{\'{\i}}nez-Mekler}}}]{Zanudo}
\bibinfo{author}{\bibfnamefont{J.~G.~T.} \bibnamefont{{Za{\~n}udo}}},
  \bibinfo{author}{\bibfnamefont{M.}~\bibnamefont{{Aldana}}}, \bibnamefont{and}
  \bibinfo{author}{\bibfnamefont{G.}~\bibnamefont{{Mart{\'{\i}}nez-Mekler}}},
  \bibinfo{journal}{ArXiv e-prints}  (\bibinfo{year}{2010}),
  \eprint{1011.3848}.

\end{thebibliography}

\end{document}